\documentclass[epj]{webofc}
\usepackage[varg]{txfonts}   
\usepackage{graphicx,wrapfig,lipsum}
\usepackage{amssymb,amsmath}
\usepackage{subeqnarray}
\usepackage{hyperref}
\newcommand{\nc}{\newcommand}   
\nc{\req}[1]{Eq.\,(\ref{#1})}    \nc{\reqp}[1]{Eq.\,(\ref{#1}) on page \pageref{#1}}     
\nc{\rf}[1]{Fig.~\ref{#1}}   \nc{\rfp}[1]{Fig.~\ref{#1} on page \pageref{#1}}     
\nc{\rt}[1]{table~\ref{#1}}   \nc{\rtp}[1]{table~\ref{#1} on page \pageref{#1}}     
\nc{\Th}{\ensuremath{T_\mathrm{H}\,}}
\nc{\pp}{\ensuremath{pp\ }}
\nc{\pA}{\ensuremath{p A\ }}
\nc{\hAA}{\ensuremath{AA\ }}
\nc{\D}{\mathrm{d}}
\nc{\E}{\mathrm{e}}
\def\beq{\begin{equation}}
\def\eeq{\end{equation}}

\wocname{EPJ Web of Conferences}
\woctitle{ICFNP2015}
\begin{document}
\title{Hagedorn legacy}
%
%

\author{Johann Rafelski}

\institute{\hspace*{-1.3mm}Department of Physics, The University of Arizona, Tucson, AZ 85721, USA}
\abstract{%
These remarks open the one-day session \lq\lq 50 years of Hagedorn\rq s Temperature and  the Statistical Bootstrap Model\rq\rq. These developments set the path at CERN towards the discovery of Quark-Gluon Plasma in the year 2000.
\centerline{\hspace*{-3cm}\includegraphics[width=0.78\columnwidth,clip]{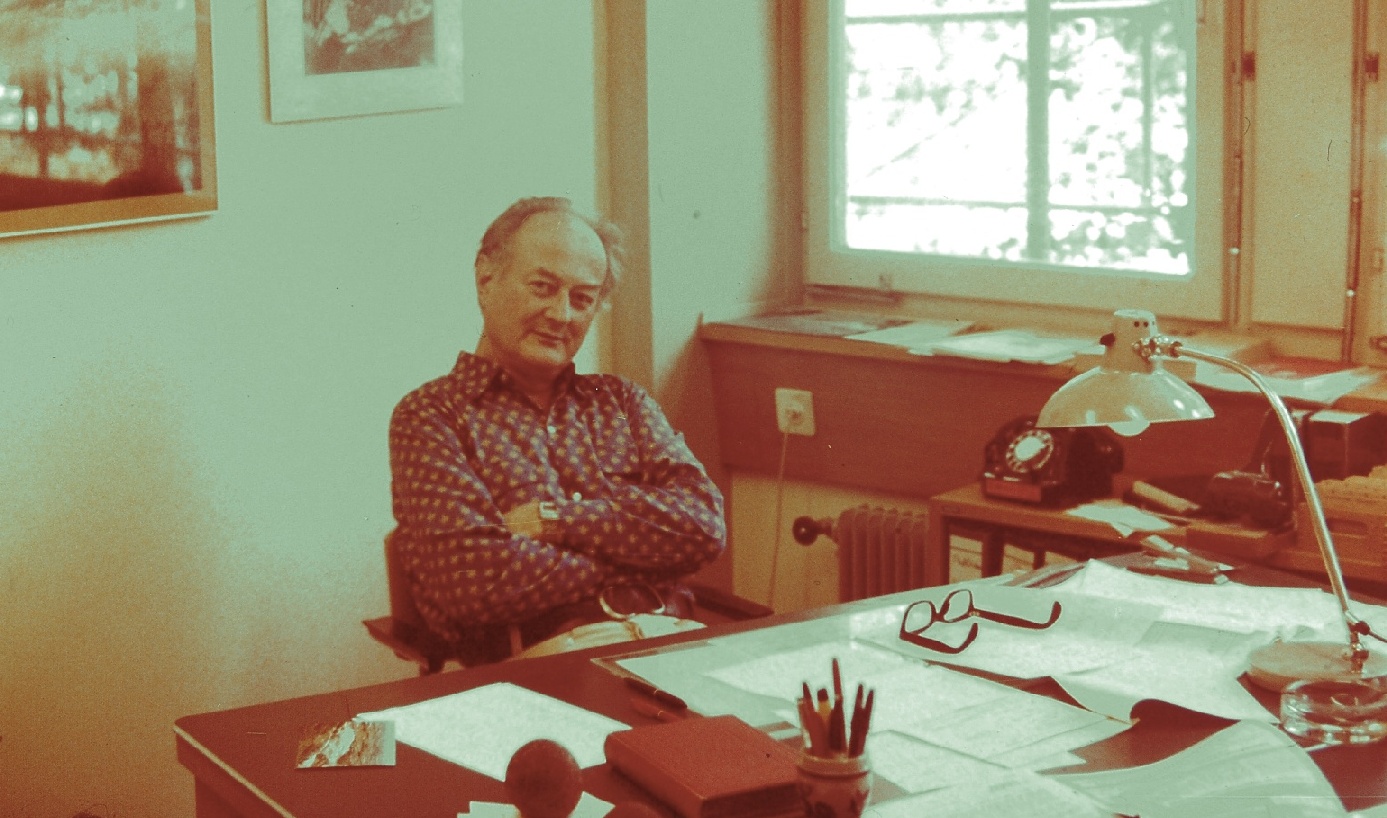}}
{\bf {\small Rolf Hagedorn in his CERN office in 1977/8, Photo: Johann Rafelski}}
}
\maketitle
%
In 1964/5 Rolf Hagedorn single handiedly had set  CERN on the path to the study of hot hadronic matter  and quark-gluon plasma. This session of the ICFNP2015  meeting has been set aside  to celebrate  50 years of Hagedorn\rq s pivotal ideas which opened to  study the  high energy density matter defining our Universe in primordial times above Hagedorn temperature \Th; that is, before the emergence of matter as we know it.   \Th\ is  the pivotal concept that gave birth to a new field of physics. Thanks to   early Hagedorn ideas,  we understand  today the different phases of strongly interacting matter, the early Universe, and we  gained a deep insight into the working of quark confinement.

\begin{figure}
\centering
\sidecaption
\includegraphics[width=0.43\columnwidth]{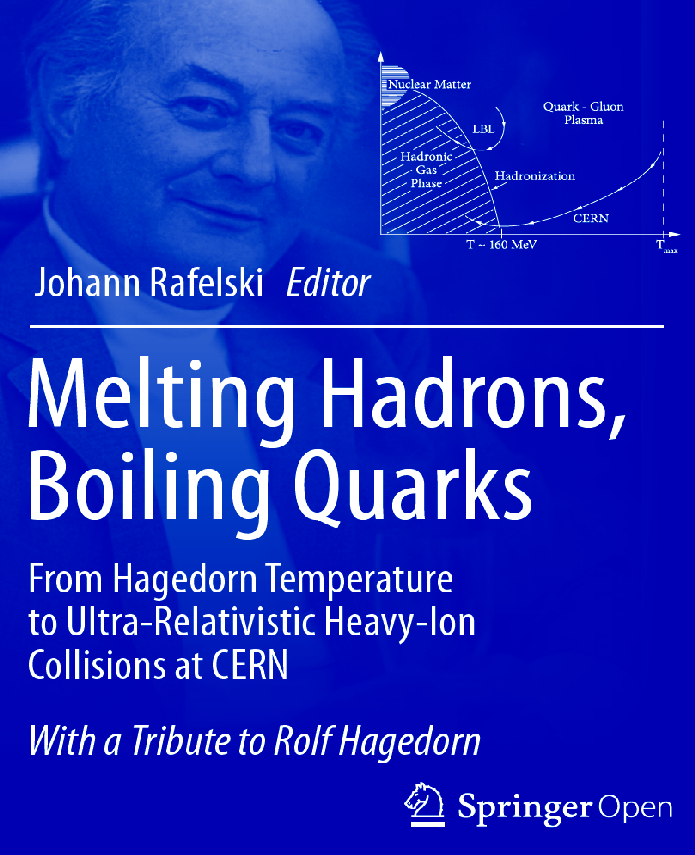}
\caption{The tale of Hagedorn temperature and its impact on today\rq s science is told in {\it Melting Hadrons, Boiling Quarks: From Hagedorn Temperature to Relativistic Heavy-Ion Collisions at CERN; With a Tribute to Rolf Hagedorn,} \url{http://www.springer.com/de/book/9783319175447},  see~\cite{Rafelski:2016hnq}.}
\label{HageCover}       
\end{figure}

Let me begin with some personal remarks:   I first met \lq Herr Hagedorn\rq\  in Winter 1975/76, at his Colloquium on the Statistical Bootstrap Model (SBM). After the lecture and some discussion that followed, I asked if I could visit him at CERN.  I arrived as a CERN-fellow in September 1977. The two most important impressions: Herr Hagedorn was an extraordinary thorough and understanding teacher; Herr Hagedorn  was available to help those who were in need.  For a more comprehensive account see the special volume~\cite{Rafelski:2016hnq} dedicated to his life and work, see figure~\ref{HageCover}.

After the vagaries of WW\,II, the path of Hagedorn to CERN was direct, as was his preparation for the work and discoveries he was to accomplish. In the early 1950s  Werner Heisenberg hired Rolf Hagedorn, who was trained by the preeminent  thermodynamics theorist Richard Becker, to work on his cosmic emulsion-star \lq hobby\rq. From the present day perspective this was the precursor research into hot hadronic matter. Heisenberg encouraged Hagedorn  to consider CERN for continuing this work. Hagedorn consequently applied on January 27, 1954  to join the future CERN laboratory, with  thermal physics and cosmic event stars as his scientific preparation. Hagedorn was hired by J.B. Adams, however, first to help  build the 1st CERN accelerator, the PS. A few years on when  J.B. Adams became director of CERN,  he assigned Hagedorn effective January 1, 1961 to a permanent position in the CERN Theory Division, charging Hagedorn with research tasks that Hagedorn dutifully executed in the following decades: \lq\lq It is intended that you devote your time partly to investigations and computing problems associated with the experimental programs of CERN, such as the use of statistical models for predicting particle production, and partly to those aspect  of theoretical physics that will enable you to keep abreast of modern developments\ldots.\rq\rq
 
Hagedorn\rq s work of 1960-64 had set  the stage for his big discoveries that followed in Winter 1964/65. The pivotal period 1964/65 was a singularly critical event in the history of modern physics as several new ideas emerged, which  are today merging into a unified view about our Universe:
\begin{itemize}
\item Hagedorn proposed the Hagedorn Temperature \Th\    and the statistical description of elementary particles, the statistical bootstrap model (SBM);
\item Gell-Man and Zweig independently recognized that strongly interacting particles have more elementary constituents, the quarks; 
\item Higgs mechanism was recognized,  opening the door for the standard model of particle physics;
\item The cosmic microwave background was discovered helping to establish the hot big-bang theory.
\end{itemize}
Akin to quarks, the Hagedorn temperature  had an uneasy birth:
Hagedorn work of 1960-1964 showed that the Fermi model produces too few pions - could this mean particles are distinguishable? Hagedorn prepared a report~\cite{Hagedorn:1964zz} that was a direct precursor of all his coming discoveries. But   he reconsidered and withdrew his preprint as new ideas were born quickly. 

Within a few more weeks Hagedorn realized   how particles can be made distinguishable: he introduced an exponential hadron spectrum where  despite large particle abundance, no particle is the same as the other. He created the bootstrap model to compute this spectrum. And, he  found that the spectrum he theoretically predicted agreed with experiment.   This was  the Hagedorn revolution of November 1964-January 1965, published as his monumental model~\cite{Hagedorn:1965st}: the SBM: Statistical Bootstrap Model.

\begin{figure}
\centering
\includegraphics[width=0.56\columnwidth]{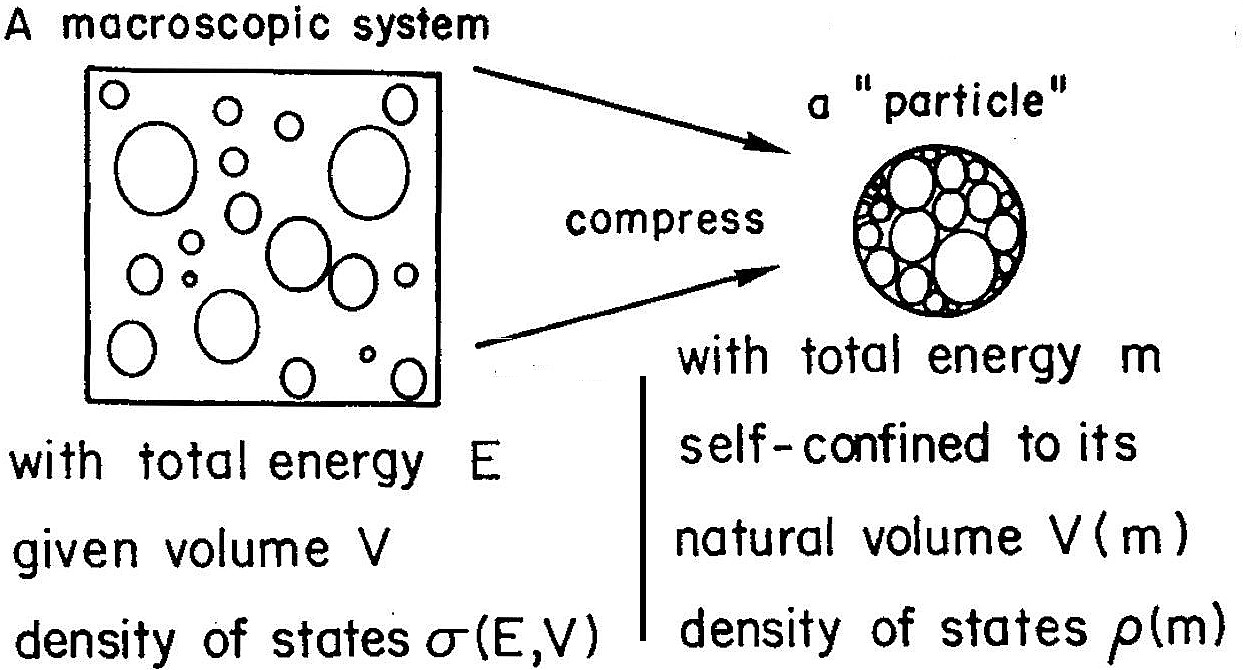}\\
\includegraphics[width=0.43\columnwidth]{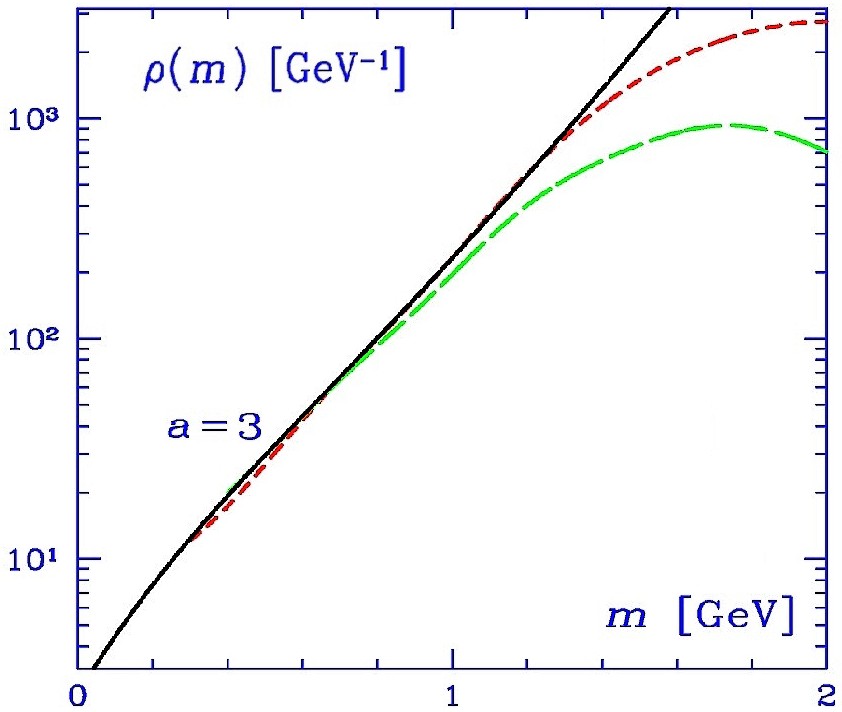} 
\includegraphics[width=0.46\columnwidth]{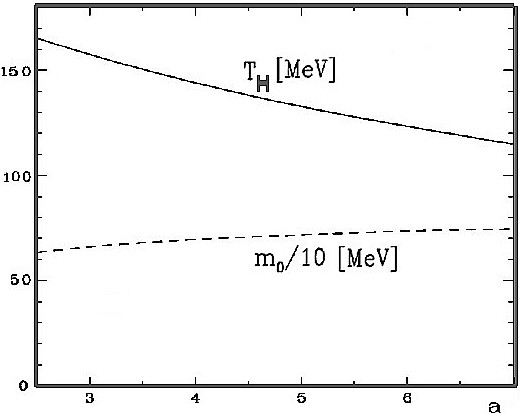} 
\caption{Top: Hagedorn\rq s picture illustration of his model prepared for review of 1984~\cite{Hagedorn:1984hz}. Bottom 
on left: The mass spectrum for $a=3$, compared to two sets of experimental data: available in early days to Hagedorn (long dashed, green) and recent vintage (short-dashed, red). On right: the parameters of \req{MassSpecEq}: \Th\ and $m_0$ as functions of the power index $a$.}
\label{MassSpecTh}       
\end{figure}

The idea of SBM was straightforward, a volume $V$ filled with particles to the limit is itself a new hadron. This is illustrated in figure~\ref{MassSpecTh}, top part. This yielded the exponential mass spectrum of hadronic states 
\begin{equation} \label{MassSpecEq}
\rho\propto \displaystyle\frac{\E^{m/\Th} }{(m_0^2+m^2)^{a/2}} \;,
\end{equation}
where the exponential slope was the Hagedorn temperature \Th. The index $a$ was at first believed shown by Hagedorn to have a value $a=2.5$.

In order to determine   \Th\ given   a limited range of experimental data available for the hadron resonance mass spectrum,  see left hand, bottom, of  figure~\ref{MassSpecTh},  Hagedorn had to use a   value of the parameter $a=2.5$. A greater value  $7\ge a\ge 3$ emerged in a more realistic versions of SBM. As we see on right in figure~\ref{MassSpecTh} the larger is $a$, the smaller is the expected value of the Hagedorn temperature. For a  discussion of this situation see Ref.\cite{Rafelski:2015cxa}. Therefore the value of \Th\ which was at first reported  to be at $\Th=164$\,MeV, was  recognized to be well below this  upper limit. The mechanism that increased value of $a$ and decreased value of \Th\ was  the canonical conservation of baryon number, and strangeness.

\begin{figure}
\centering
\includegraphics[width=0.44\columnwidth]{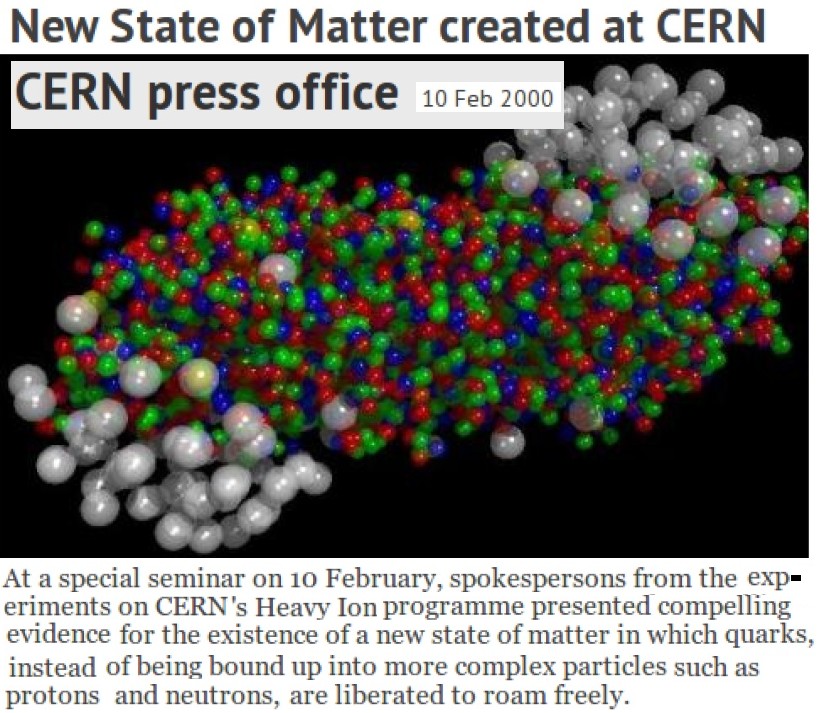} 
\includegraphics[width=0.49\columnwidth]{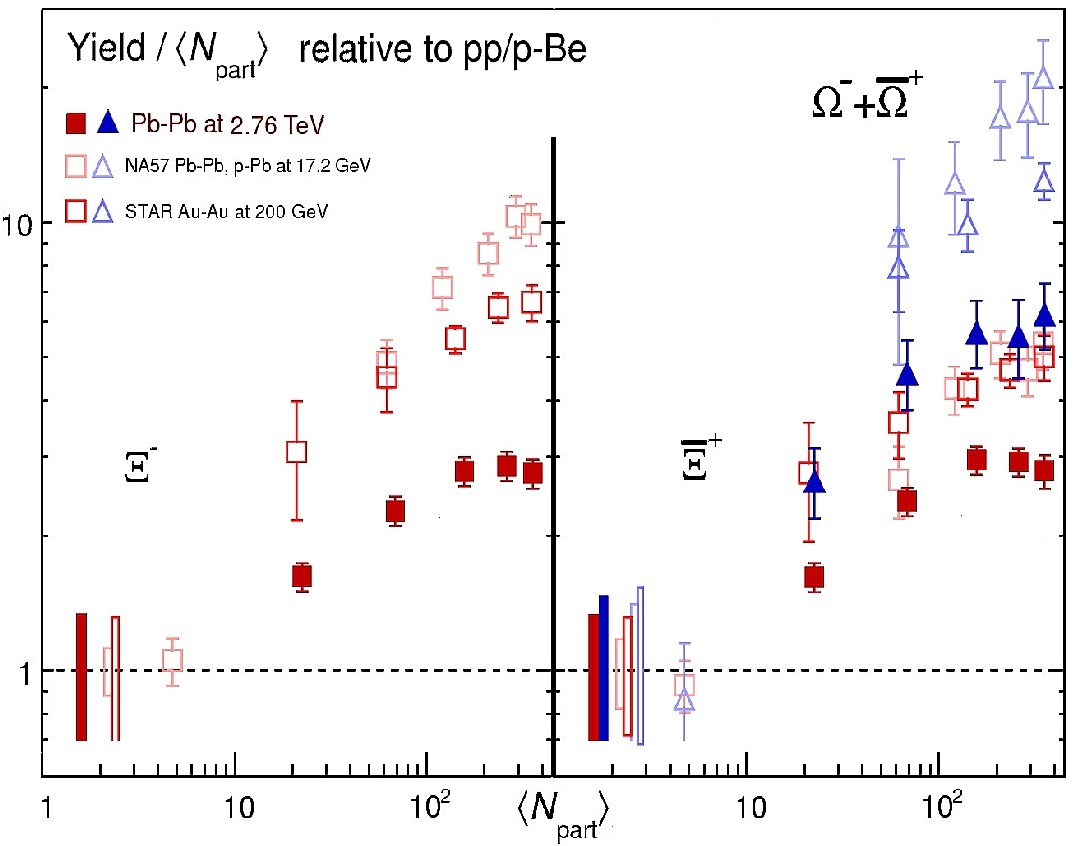} 
\caption{On left: CERN announces the discovery of quark-gluon plasma \url{press.web.cern.ch/press-releases/2000/02/new-state-matter-created-cern}; On right: the 2014 summary of strange antibaryon signature of QGP, adapted from a CERN-LHC-Alice collaboration report~\cite{ABELEV:2013zaa}.}
\label{QGPdisc}       
\end{figure}

The work of Hagedorn has attracted interest by many researchers; some are today here. They  followed in Hagedorn\rq s footsteps, often working at CERN. We were   using  statistical methods in strong interaction physics. After 1977 this effort turned into the exploration of  hot hadronic matter created in nuclear collisions. This effort had created at CERN a hot spot in a new research field that was not fully appreciated everywhere at first. It was this development which made it  natural that in late 1970s and early 1980s CERN became the host to a new experimental research program into the physics of relativistic heavy ion collisions. Ultimately, this led to the announcement of  the discovery of quark-gluon plasma  at CERN in early 2000, see figure~\ref{QGPdisc} on left. Among the prominent signatures was the  strange antibaryon production enhancement, the largest and predicted effect of quark deconfinement  in hot hadron matter, shown on the right in figure~\ref{QGPdisc}, see Refs.\,\cite{Rafelski:2016hnq,Rafelski:2015cxa}. 

\end{document}